\documentclass[a4paper, 10pt, conference]{ieeeconf}      

\IEEEoverridecommandlockouts                              

\usepackage{graphics} 
\usepackage{epsfig} 
\usepackage{booktabs}
\usepackage{tikz}

\newcommand\copyrighttext{%
  \footnotesize \textcopyright 2023 IEEE. Personal use of this material is permitted. Permission from IEEE must be obtained for all other uses, in any current or future media, including reprinting/republishing this material for advertising or promotional purposes, creating new collective works, for resale or redistribution to servers or lists, or reuse of any copyrighted component of this work in other works.}
\newcommand\copyrightnotice{%
\begin{tikzpicture}[remember picture,overlay]
\node[anchor=south,yshift=10pt] at (current page.south) {\fbox{\parbox{\dimexpr\textwidth-\fboxsep-\fboxrule\relax}{\copyrighttext}}};
\end{tikzpicture}%
}

\title{\LARGE \bf
Unified Occupancy on a Public Transport Network through Combination of AFC and APC Data
}

\author{Amir Dib, Noëlie Cherrier$^*$, Martin Graive, Baptiste Rérolle and Eglantine Schmitt$^{1}$
\thanks{$^*$Corresponding author}%
\thanks{$^{1}$Authors are with Citio, 54 Quai de la Rapée, 75012 Paris France.
        {\tt\small \{firstname\}.\{lastname\}@cit.io}}%
}

\begin{document}

\maketitle
\copyrightnotice
\thispagestyle{empty}
\pagestyle{empty}

\begin{abstract}
In a transport network, the onboard occupancy is key for gaining insights into travelers' habits and adjusting the offer. Traditionally, operators have relied on field studies to evaluate ridership of a typical workday. However, automated fare collection (AFC) and automatic passenger counting (APC) data, which provide complete temporal coverage, are often available but underexploited. It should be noted, however, that each data source comes with its own biases: AFC data may not account for fraud, while not all vehicles are equipped with APC systems.

This paper introduces the unified occupancy method, a geostatistical model to extrapolate occupancy to every course of a public transportation network by combining AFC and APC data with partial coverage. Unified occupancy completes missing APC information for courses on lines where other courses have APC measures, as well as for courses on lines where no APC data is available at all. The accuracy of this method is evaluated on real data from several public transportation networks in France.
\end{abstract}

\section{INTRODUCTION}

Both public transportation operators and authorities organizing the mobility on their territory (e.g. city halls) require an understanding of how their transportation networks are utilized. They can adapt accordingly their transport offerings, including vehicle capacities, service frequency, and schedules, by examining the temporal fluctuations in passenger flows. Moreover, they can leverage this information to gain in-depth knowledge of commuting patterns, such as public transport usage in relation to geographic areas.

In all these scenarios, a comprehensive understanding of the network's ridership is critical. Historically, field surveys have been prevalent and continue to be widely used today: personnel conduct on-site questioning of commuters about their current trip's origin, destination, purpose, general habits, etc. An alternative technique is to count individuals entering and exiting vehicles at each stop. These surveys typically represent a standard business day, given that a statistical adjustment is conducted to account for the survey sample's characteristics. However, they do not account for exceptional events and tend to underestimate fair evasion \cite{MeasuringandControlling}. In addition, they do not permit to quickly validate the efficiency of anti-fraud measures. With the advent of advanced information systems, data offering insights into the network's ridership is readily available, but underexploited. For instance, ticketing data is now often automated and consolidated into databases but rarely provides alighting information, except in the few networks requiring validation at both entry and exit points. Automated Fare Collection (AFC) data also omits fraud ridership, paper tickets, cellphone validations, and system failures \cite{MeasuringandControlling}, thereby not fully representing network occupancy. Conversely, Automated Passenger Counting (APC) systems are increasingly being adopted by operators, utilizing technologies such as optical sensors, pressure sensors, computer vision, and Wi-Fi or Bluetooth-based systems, each with their advantages and drawbacks \cite{grgurevic2022review,comparisonWLAN}. These systems can provide a count of individuals entering and leaving the vehicle (for instance with sensors at the doors), or directly measure occupancy (using cameras, or weight sensors). However, they remain optional for network operations and are often viewed as expendable costs by the operators. A common practice is to equip part of the vehicle fleet and rotate the equipped vehicles to gradually cover the entire network.

This study aims at optimizing the use of all available information, {i.e.}, integrating ticketing data from AFC systems and counting data from APC systems to obtain comprehensive ridership information for all courses in any temporal range of a public transportation network. We combine the strengths of both data sources, leveraging the extensive coverage of ticketing data and the comprehensiveness of APC data, to derive a fraud model over the entire network. This fraud model enriches the ticketing information to produce what we term the \textit{unified occupancy}. Consequently, the transport operator receives detailed information at the spatial and temporal levels, as well as a distinction between regular and fraudulent ridership. The proposed method is evaluated on real data from four public transportation networks to determine its accuracy and interpret its results.

Section~\ref{sec:related_work} provides an overview of the related work in this field, including a discussion on occupancy prediction. Section~\ref{sec:proposed_method} presents our approach, which includes data preprocessing and fraud modeling. Finally, comprehensive experiments on real-world data are conducted in Section~\ref{sec:experiments}.

\section{RELATED WORK}
\label{sec:related_work}

There is abundant scientific literature on several aspects of our subject, though none providing an unified occupancy model combining AFC and APC data to our knowledge. Work has been conducted on the relevance of field surveys and the causes of fraud. The usage of automatically collected data is studied in a variety of forms, in particular to extrapolate or to predict fraud and/or occupancy.

Field surveys and interviews are the main means of quantifying and identifying the causes of fraud on a network \cite{delbosc2016four,cantillo2022fare}. Several motivations stand out: accidental evasion, ideological fraud, and economically-compelled evasion \cite{barabino2020fare}. However it has also been shown that field surveys and interviews are subject to numerous biases: for instance on crowded lines or large vehicles, inspection cannot be exhaustive and the fraud rates are usually underestimated \cite{egu2020can}; likewise, the behaviour of the fare evaders changes upon seeing the inspectors which leads to an underestimation of the fraud rates \cite{delbosc2019people}. Still, a multiplicative factor can be derived from these surveys and used to correct the occupancy resulting from ticketing data. This factor is often given for the whole network or at best for each line and therefore assumes fraud is evenly distributed over the line/network \cite{yap2020crowding}. 

The usage of both APC and AFC data to model fraud is established in the scientific literature \cite{pourmonet2015vers,egu2020can}. The ratio of validations over boarding counts is used and interpreted as the ``rate of fare non-validation''. This value is dubbed by the authors a promising alternative to the assessments of field surveys. However the gathered information is station-wise, not course-wise, and in the event of an unequipped line, no fraud can be inferred. In this work, we propose a method to get occupancy at the course level for all courses in the network.

Recently, Roncoli \textit{et al.} made use of incomplete APC data to extrapolate occupancy in Nantes, France \cite{roncoli2023estimating}. They developed a Kalman filter based method to get estimates of boardings and alighting rates per station. The downside of this method is that it is not robust to volatility in passenger flows when the system is not observable.

The goal of this work is to fill in the missing occupancy data in past courses, while predictive approaches focus on assigning occupancy predictions to courses in the future. Therefore, we present a few works focusing on occupancy forecasting.

It is possible to predict a subset of fraud based solely on AFC data, as is done in \cite{sanchezMartinezEstimatingNoninteraction}. The authors devise a stochastic model with priors computed on smart card holders' supposed non-interaction on their frequent commutes. However, this approach is inadequate to predict fraud by occasional travelers. It is also vulnerable to spikes in traveler numbers, for instance in connection with citywide events.

Alternatively, occupancy prediction making use of APC systems has been extensively addressed by the scientific community \cite{pasini:hal-02278238}. Auto-regressive models for time-series forecasting have been investigated \cite{xue2015short,TengChen}, as well as lasso estimators \cite{jenelius2019metro,jenelius2019data} and neural networks approaches \cite{talusan2022designing,Toque2016,TengChen}. The method that seems to prevail today is the use of random forest regressions \cite{Vial_Gazeau_2020,arabghalizi2019full,WOOD2022,Toque2017}. All these approaches, including the simplest ones, are equal in regard to the shortcomings of the APC data they use. This is why our proposal is to use AFC data in addition.

Aggregation between ticketing and counting data exist for reconstructing Origin/Destination (O/D) matrices, as in \cite{gordon2018estimation}, where the APC data is used to scale the O/D matrix extracted from iterative proportional fitting on AFC data. In \cite{amblard2023bayesian}, a Bayesian Markov model for trip inference with APC and AFC priors is proposed. Milkovits \cite{milkovits2008modeling} uses AFC to correct APC under-counting when predicting bus stop dwell times. The literature, to our knowledge, does not take into account both the ticketing data and the counting cell measures to infer occupancy.

\section{FRAUD MODELING TO UNIFY OCCUPANCY FROM TICKETING DATA}
\label{sec:proposed_method}

\begin{figure}
  \centering
  \includegraphics
  [width=\linewidth]{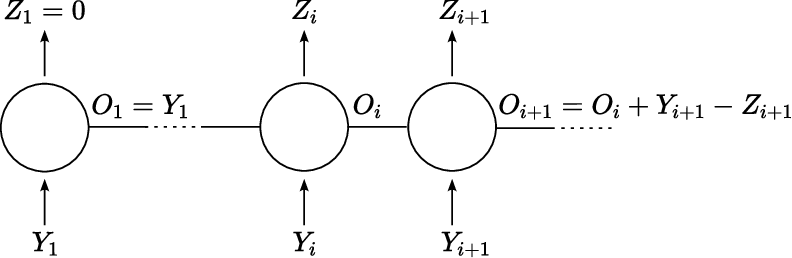}
  \caption{Boardings ($Y_i$), alightings ($Z_i$) and occupancies ($O_i$) over a network line.}
  \label{fig:stations_occupancy}
\end{figure}

Occupancy is computed from a measure of boardings and alightings at each station through the following formula:
\begin{equation}
\label{eq:occupancy_from_boardings_alightings}
    O_i = \sum_{i=1}^{N} Y_i - Z_i,
\end{equation}
with $i$ the position of the station on the line, $Y_i$ (resp. $Z_i$) the number of boarding (resp. alighting) passengers, and $O_i$ the occupancy between stations $i$ and $i+1$. Since no passenger can alight at the first station or board at the last station, $Z_1 = Y_N = O_N = 0$ ($N$ being the number of stations).
Fig.~\ref{fig:stations_occupancy} illustrates the basics of occupancy, boardings and alightings on a network line.

The total occupancy $O_i$ breaks down into ticketing occupancy $O_i^V$ (including passengers who validated their ticket) and fraudsters occupancy $O_i^F$, so that $O_i = O_i^V + O_i^F$.

The study leverages two sources of data. The first source is boarding ticketing data, providing comprehensive knowledge of $Y_i^V$ for all stations within the network. The second source is counting cells data, which may offer either occupancy or boardings and alightings information, depending on the technology. In this study, the proposed method presumes that partial knowledge of $O_i$ is available across the network, meaning that counting cells occupancy is known for a subset of courses, which does not necessarily cover all lines of the network. If required, occupancy can be calculated from boardings and alightings using~(\ref{eq:occupancy_from_boardings_alightings}).

The goal of this study is to reconstruct the actual unobserved occupancy on the vehicle for each station of every course, {i.e.}, obtaining a complete knowledge of $O_i$ given: $O_i$ on a subset of courses, and $Y_i^V$ on all courses.

\subsection{General principle}

The proposed method involves modeling fraud at the station level. Indeed, the integration of ticketing data and counting cells data, where available, provides insights into the spatial distribution of fraud. The principle is to establish a station-level rate $R_i$ such that
\begin{equation}
\widetilde{O}_i^F = O_i^V R_i,\quad \widetilde{O}_i = O_i^V + \widetilde{O}_i^F.
\label{eq:unification}
\end{equation}

In essence, a fraud rate is aggregated at the station level from all the known counting cells and ticketing information. Subsequently, inference for an unknown occupancy involves applying this fraud rate to ticketing occupancy to estimate the total occupancy including fraudsters.

\begin{figure*}
  \centering
  \includegraphics[width=\textwidth]{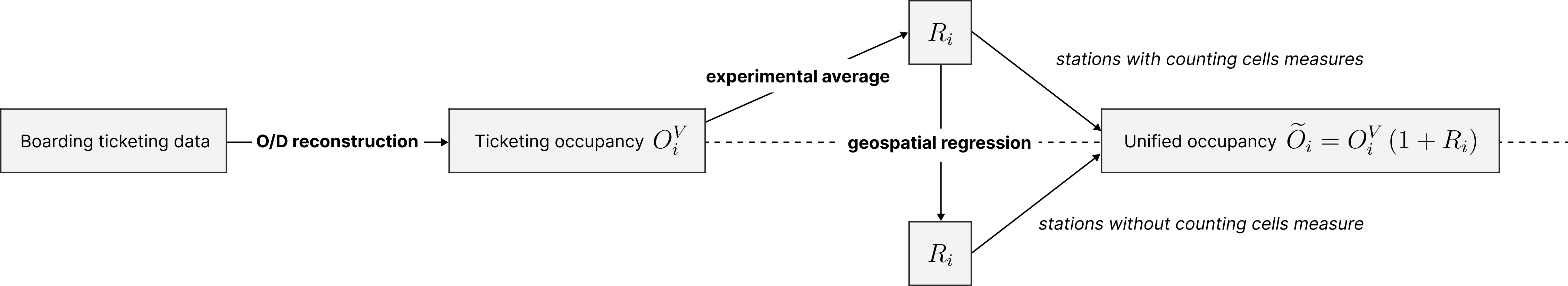}
  \caption{Summary of unified occupancy method. All courses go through O/D reconstruction, then courses with APC data are used to compute average $R_i$ for a subset of the stations. Geospatial regression provides $R_i$ for the stations without counting cells measures. In the end, the $R_i$ are used to compute the unified occupancy for courses without APC data.}
  \label{fig:method_summary}
\end{figure*}

The method consists of five steps outlined in Fig.~\ref{fig:method_summary}. Since AFC data is available for all courses, the first step is to apply O/D reconstruction (Section~\ref{sec:preprocessing_ticketing}). Courses with APC data (Section~\ref{sec:preprocessing_cc}) are then used to calculate individual fraud rates from counting cells occupancy $O_i$ and ticketing occupancy $O_i^V$. For stations with counting cells measures (top half of Fig.~\ref{fig:method_summary}), fraud rates $R_i$ are averaged (Section~\ref{sec:fraud_with_measures}). Subsequently, a geospatial regression allows for the computation of $R_i$ values across the entire network, including stations without counting cells measures (bottom half of Fig.~\ref{fig:method_summary}, see Section~\ref{sec:fraud_without_measures} for details). Finally, courses without counting cells measures utilize the computed $R_i$ at each of their stations to compute unified occupancy as in~(\ref{eq:unification}).

It is worth noting that the fraudster rate at each station $R_i$ differs according to the type of data available on a station: some stations in the network may never be served by courses with counting cells data. Thus, a distinct treatment is applied. The following subsections elaborate on how data is processed and the fraudster rate modeled.

\subsection{Data preprocessing}

\subsubsection{Ticketing data}
\label{sec:preprocessing_ticketing}

The objective here is to estimate $O_i^V$ from the complete knowledge of $Y_i^V$, which requires the complementary availability of $Z_i^V$. While only few networks have both boarding and alighting ticketing systems, providing straightforward computation of $O_i^V$ through~(\ref{eq:occupancy_from_boardings_alightings}), most systems record only boarding validations, thus making computation of occupancies more challenging.

Nevertheless, there is rich literature on O/D reconstruction, which involves identifying the alighting station associated with each boarding validation \cite{yan2019alighting,assemi2020improving,luca2021survey}. Trip chaining \cite{trepanier2007individual,li2018smart,hussain2021transit}, the most traditional method, allows for the linking of consecutive validations of a single user using their smart-card data. In this study, we adopt the Bayesian approach of \cite{amblard2023bayesian} that combines counting cells and ticketing data to reconstruct trips for non-smart-card users. Ultimately, this reconstruction provides a list of boarding validations and alighting stations, which can be combined using~(\ref{eq:occupancy_from_boardings_alightings}) to obtain ticketing occupancies $O_i^V$ for all stations on all courses.

\subsubsection{Counting cells data}
\label{sec:preprocessing_cc}

It should be mentioned that due to sensor inaccuracies and biases we only have access to $O_i^C$, which is a noisy version of the true unobserved occupancy $O_i$. This work assumes $O_i^C = O_i$, but solutions exist to denoise counting cells data and enforce passenger flow conservation \cite{kikuchi2006method,de2014adjustment,yin2017}.

\subsection{Fraud computation for stations with some counting cells measures}
\label{sec:fraud_with_measures}

From the reconstructed $O_i^V$ and measured $O_i$, the objective is to compute a fraud rate $R_i$ for all stations in the network. First, the proposed method focuses on stations in the network with at least one course stopping at station $i$ with $O_i$ information derived from counting cells data. Other courses without counting cell data stopping at these stations will be attributed a unified occupancy estimation $\widetilde{O}_i$ from the averaged fraud rates of equipped courses on the same route.

For these stations, $R_i$ is averaged over all courses that have counting cells data:
\begin{equation}
    R_i = \frac{1}{\left|\mathcal{C}_{i,l}\right|} \sum_{c \in \mathcal{C}_{i,l}} \frac{O_{i,c} - O_{i,c}^V}{O_{i,c}^V}
\end{equation}
with $\mathcal{C}_{i,l}$ the set of courses with counting cells on line $l$ stopping at station $i$, $O_{i,c}$ the measured counting cells occupancy by course $c$ at station $i$ and $O_{i,c}^V$ the reconstructed ticketing occupancy for course $c$ at station $i$.

\subsection{Fraud computation for stations with no counting cells measure}
\label{sec:fraud_without_measures}

After averaging the fraud rate for stations that have counting cells data, a few stations or even entire lines of the network may have no $R_i$ estimation because no course equipped with counting cells stopped there and measured ridership. For these stations, a geospatial regression model is inferred, permitting as a bonus to get a smooth estimate of the fraud levels over the entire territory.

The training dataset consists in the already inferred pairs of stations $i$ and fraud rates $R_i$ with their longitudes and latitudes. The objective is to derive $R_i$ at all points in the network, and in particular at stations without counting cells measures. 

The selected fitting method is ordinary kriging, {i.e.} Gaussian process regression, since the field of fraud rate values should be spatially correlated according to domain knowledge. Kriging derives the field value at any unknown point based on the known observed values at other points (in this case, the stations with already inferred $R_i$). The value for a new point is equal to a weighted average of the values at the known points, the weights being inferred from a provided covariance model. In this instance, an exponential model is fitted to the empirical variogram of the training dataset to obtain the covariance model for kriging. To avoid aberrant values, the $R_i$ in the training dataset are clipped to 1, namely considering that fraud cannot exceed 50\% of total ridership.

Ultimately, this fitted geospatial regression model provides not only the missing fraud rates $R_i$ at stations without counting cells measures, but at every point of the network. The scalar fields of fraud across a network in France are presented and discussed in Section~\ref{subsec:fraud_maps_interpretation}.

\section{EXPERIMENTS AND RESULTS}
\label{sec:experiments}

This section is devoted to the evaluation of the unified occupancy. First, we outline the evaluation framework and test our approach on real-world networks. Several examples of courses are presented with careful examination of the effect of sparse counting cells coverage. Finally, an interpretation of fraud maps is developed.

\subsection{Experimental setup}
\label{sec:experimental_setup}

Our experiments are performed on one month of real data (AFC and APC data) from four networks of medium-size cities in France, one of them being anonymized at its demand (noted Network A in the following).

When evaluating the performance of the unification, the main difficulty lies in the fact that there is no available ground truth. To circumvent this limitation, the counting cell measures are used as a reference.

The unification algorithm is subdivided into two models: the mean fraud rate model that is used on stations with APC data, and the geospatial regression for stations without APC. 
To evaluate the relevance of the mean fraud rate model, the fraud rates of stations with APC data are computed for each line with 30\% of counting cells measures randomly removed from the training data. The discrepancy between the unified occupancies $\widetilde O$ resulting from these fraud rates and the ``true'' occupancy $O$ is quantified using the weighted mean absolute percentage error (wMAPE), defined as
\begin{equation}
    \mathrm{wMAPE}(O,\widetilde O) = \frac{\sum_{i=1}^n |O_i-\widetilde O_i|}{\sum_{i=1}^n|O_i|},
\end{equation}
that can be interpreted as the MAPE weighted by the true value.

In the same fashion, the geospatial regression is evaluated by taking in turns a line with counting cell measures and comparing these values with the ones obtained through geospatial regression fitted on the data of the other lines. It should be noted that the performance of geospatial regression on lines that are not equipped with APC systems cannot be evaluated by this means.

\subsection{Performance evaluation}
\label{sec:performance}

In Table~\ref{tab:prediction}, our model is compared on four networks to a baseline, namely the contextual average model. As defined in \cite{pasini:hal-02278238}, the contextual average consists in averaging the known occupancy of vehicles that have similar features. We feature-engineered the following: time of the day (rounded to the quarter-hour), day of the week, stop, line and direction. The average is done on three months of data. Note that this baseline is unable to predict the occupancy on lines that do not have counts. 

In Angers for instance, the average occupancy being 25, a wMAPE of 20\% translates to an error of 5 passengers in absolute value. This comparison highlights that both the mean fraud rate model and the geospatial regression model outperform the baseline in all networks. It is likely that using the ticketing data in our models accounts for this improvement. Indeed, the validations are a reliable source of knowledge at the course level; they allow for a better treatment of outliers.

The geospatial model features the lowest errors but is normally designed for stations where no APC data was recorded. It is probable that the geospatial model smooths out the possibly aberrant fraud rates computed by the mean fraud rate model. Indeed, the average fraud rate computed at each station can be distorted by a few extreme values. Maybe utilizing the geospatial model for all stations on the network after fitting it on stations having a mean fraud rate available would yield more robust results.

\begin{table}
    \centering
    \caption{wMAPE (in \%) of different occupancy reconstruction methods}
    \label{tab:prediction}
    \begin{tabular}{lccc}
        \toprule
         Network & Contextual average & Mean fraud rate & Geospatial \\
         \midrule
         Angers & 42.7 & 19.2 & \textbf{14.0} \\  
         Nevers
         & 56.4 & 26.9 & \textbf{25.0} \\
         Brest
         & 61.8 & 46.0 & \textbf{33.3} \\
         Network A
         & 50.0 & 36.9 & \textbf{27.9} \\
         \bottomrule
    \end{tabular}
\end{table}

\begin{table}
    \centering
    \caption{Reconstruction per line in Angers and Nevers} 
    \label{tab:angers}
    \begin{tabular}{llc}
        \toprule
        City & Line & wMAPE (\%)  \\ \midrule
        Angers & 12d & 37.5 \\
        & 3s & 39.1 \\
        & 02 & 14.5 \\
        & 04 & 17.4 \\
        \midrule
        Nevers & 04 & 28.0 \\
        & 03 & 28.9 \\
        & 13 & 20.3 \\
        & 12 & 19.8 \\
          \bottomrule
    \end{tabular}
\end{table}

Table~\ref{tab:angers} displays the wMAPE per line in Angers and in Nevers, as per the performance evaluation framework presented in Section~\ref{sec:experimental_setup}. Four lines ranked by increasing mean occupancy were chosen per network.

Angers' lines 3s and 12d and Nevers' lines 03 and 04 have a low average occupancy, while lines 02 and 04 from the Angers network and 12 and 13 in Nevers are known to be heavily loaded. 
The relative error decreases in proportion to the number of passengers on the line. Therefore our model performs best on busy lines. However, the average absolute error is somewhat constant and equals $4.5$ passengers in Angers and $1.9$ passengers in Nevers.

\subsection{Real courses examples}

\begin{figure}
    \centering
    \includegraphics[width=\columnwidth]{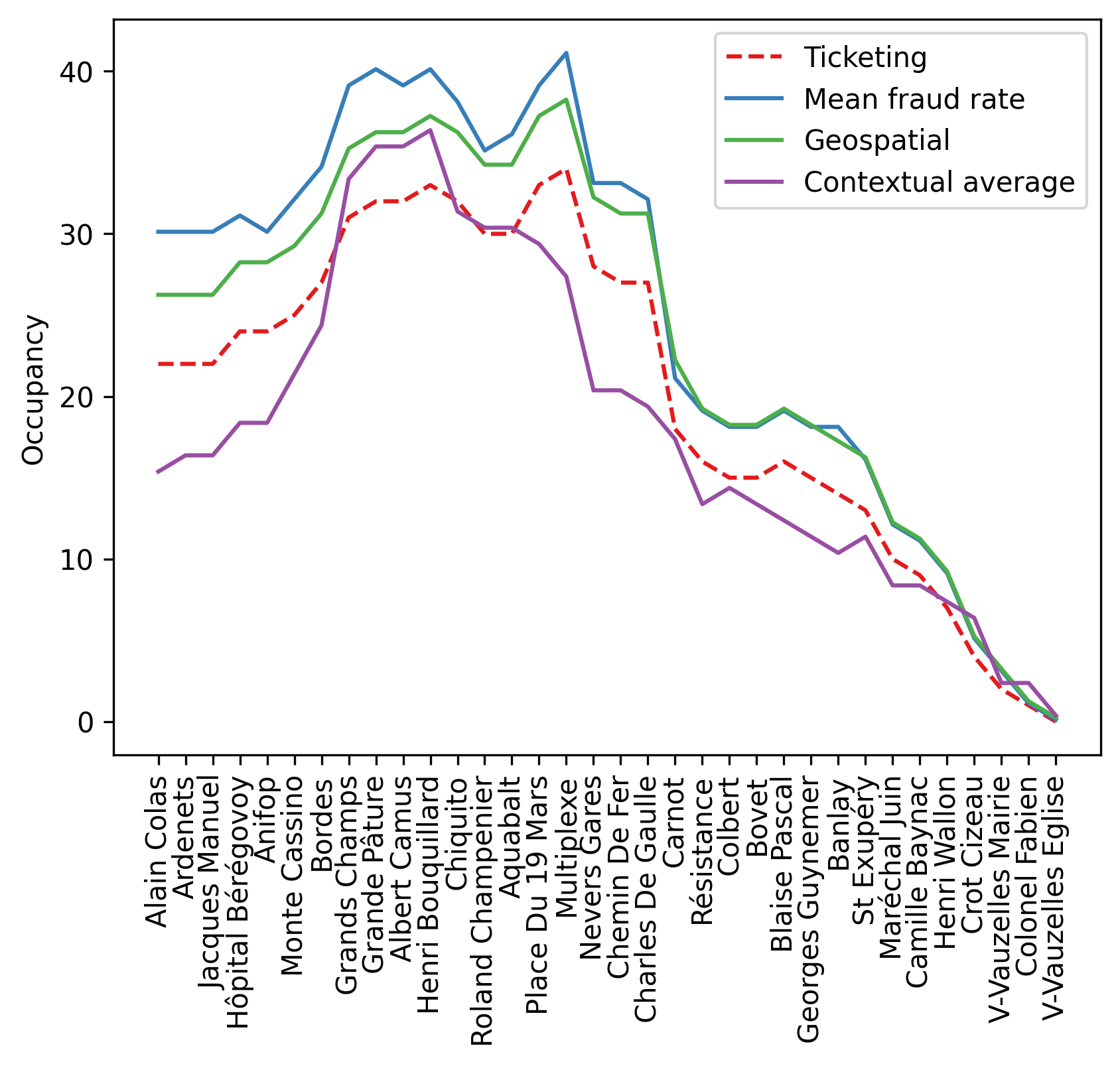}
    \caption{Occupancies of a course in Nevers in function of the serviced stops.}
    \label{fig:example_course_1}
\end{figure}

\begin{figure}
    \centering
    \includegraphics[width=\columnwidth]{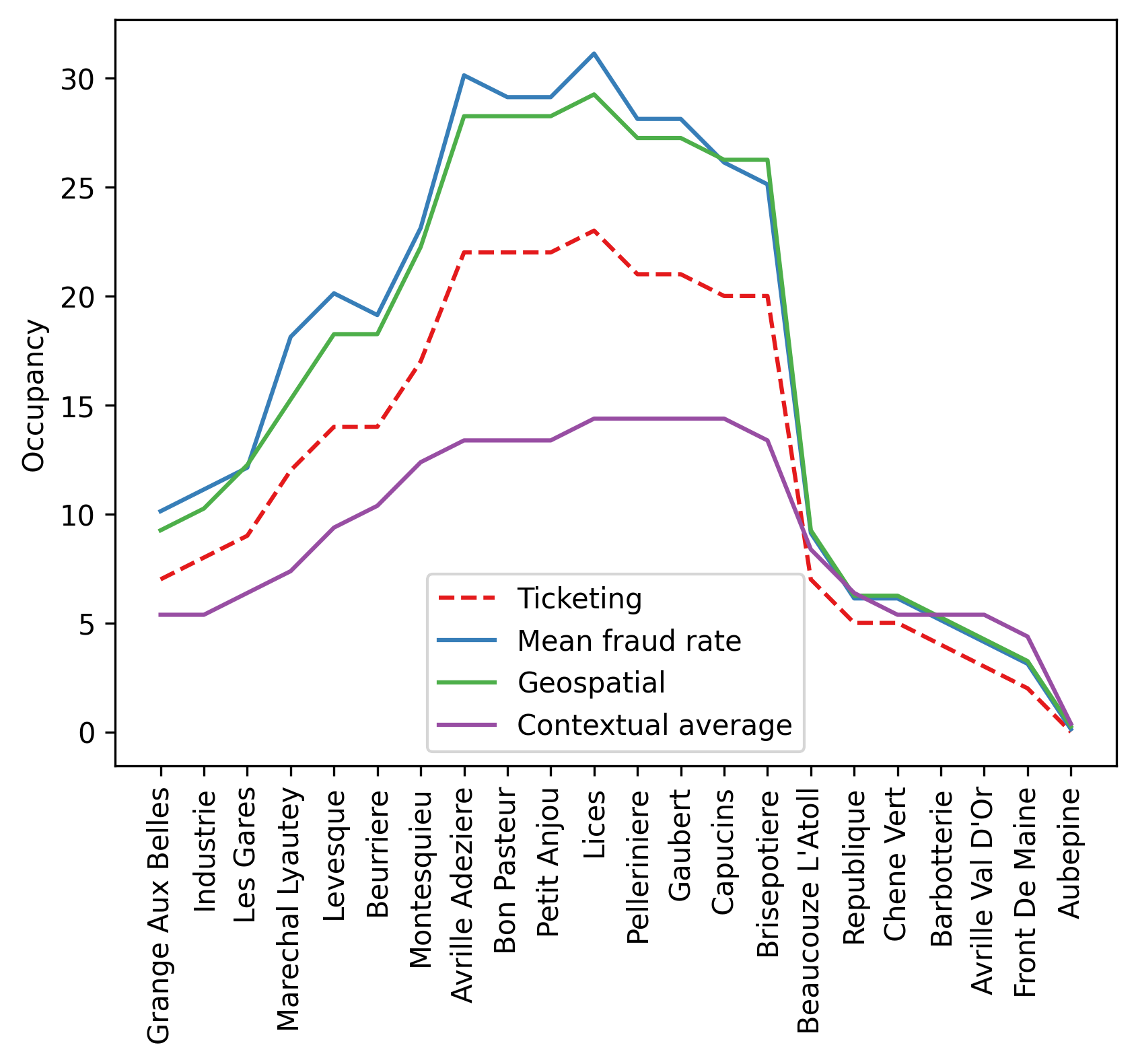}
    \caption{Occupancies of another course in Nevers in function of the serviced stops.}
    \label{fig:example_course_2}
\end{figure}

To further demonstrate the relevance of the proposed approach, a few courses are illustrated and discussed here. 
Fig.~\ref{fig:example_course_1} and~\ref{fig:example_course_2} show the different occupancies of a vehicle during its course. The ticketing occupancy (dashed line) serves as a reference. 

The occupancies reconstructed via the geospatial and mean fraud rate models stay over the occupancy reconstructed from ticketing, which makes sense because the fraudsters are added to the ticketing data. On the other hand, the extrapolated occupancy from the contextual average is irrelevant as it is lower than the ticketing occupancy in both Fig.~\ref{fig:example_course_1} and Fig.~\ref{fig:example_course_2}, even though those passengers are actually in the vehicle with high certainty. 
Moreover our models' occupancies profiles seem more consistent as the occupancy peaks are the same as for the ticketing occupancy.

On Fig.~\ref{fig:example_course_2}, we can see that the geospatial model smooths the fraud rates, as its curve follows the ticketing occupancy more than the mean fraud rate occupancy does. The ticketing occupancy is constant while the mean fraud rate occupancy is not on several segments, for example between stations Léon Blum and Centre Expositions, or between Chemin De Fer and Charles De Gaulle, which translates a fraud rate variation between adjacent stations. The fraud rate actually decreases during the course, going from the suburbs to the center of Nevers. This supports the choice of modeling a fraud rate at the station level, unlike most field surveys providing a fraud rate at the line level \cite{yap2020crowding}.

\subsection{Impact of counting cells coverage}
\label{sec:cc_coverage}

Coverage rates seldom exceed 50\% of vehicles of a line, this percentage being even lower for some routes. Thus, it is important for our model to be able to reconstruct occupancy accurately even when few vehicles have APC data.

To evaluate this limitation, a fully equipped line of the Angers network was selected to experiment on. About a hundred distinct courses across one month served as training data. 
The proposed algorithm was run multiple times on one month of data (about a hundred distinct courses), removing the counting cells data from a single course at each iteration and computing the error between reconstructed unified occupancy and original deleted APC occupancy.

\begin{figure}
    \centering
    \includegraphics[width=\columnwidth]{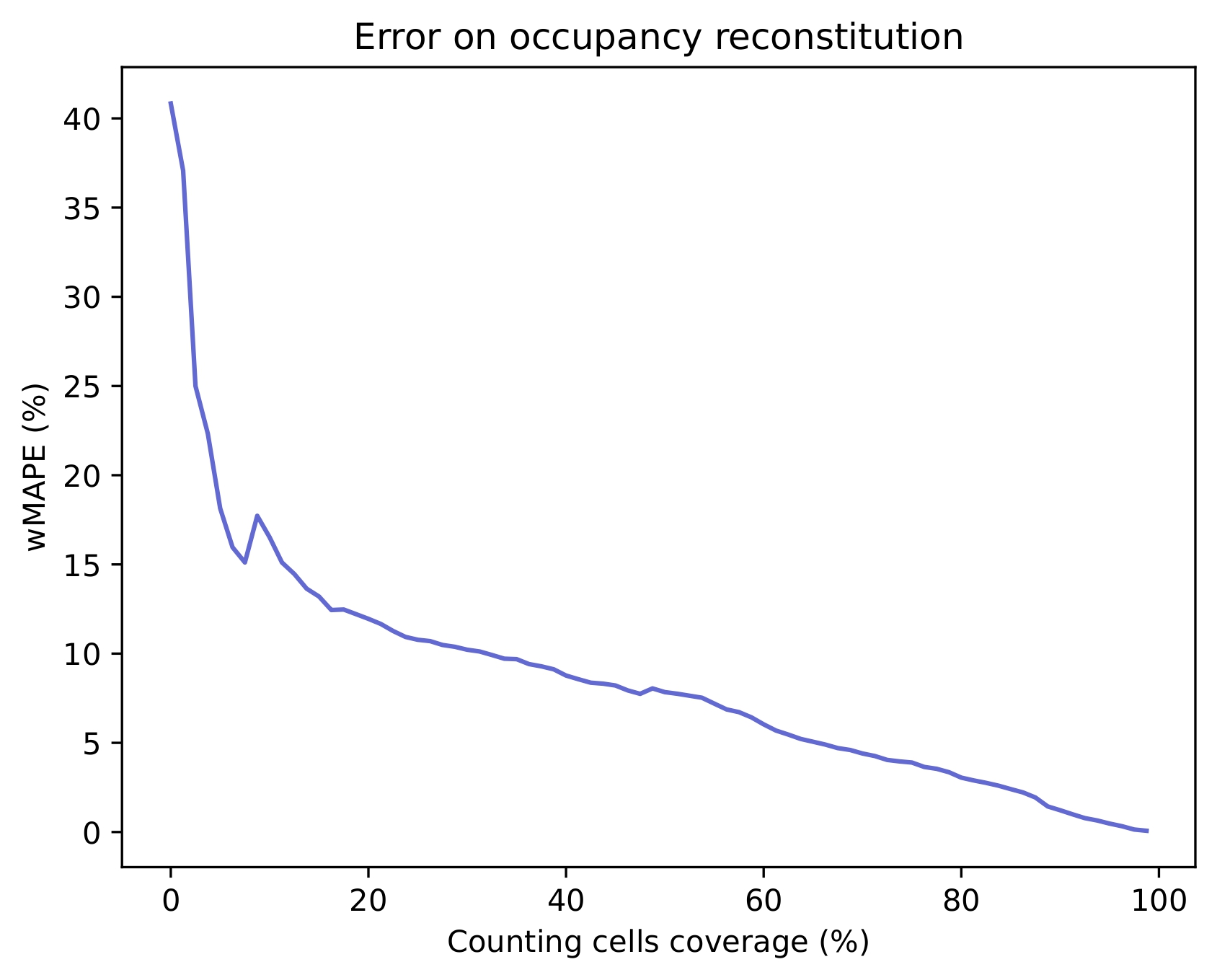} 
    \caption{Performance of occupancy reconstruction on one line initially fully equipped and of which counting cells were sequentially removed.}
    \label{fig:error_coverage}
\end{figure}

Fig.~\ref{fig:error_coverage} shows the evolution of the average error (using the wMAPE metric) on occupancy as a function of the percentage of courses with APC data. As expected, the errors diminish with increasing coverage, with an elbow at around 10\% coverage. Note that even with coverage as low as 10\%, errors are reasonable with a maximum deviation of 15\% from the true counts.

\subsection{Fraud map interpretation}
\label{subsec:fraud_maps_interpretation}

\begin{figure}
    \centering
    \includegraphics[width=\columnwidth]{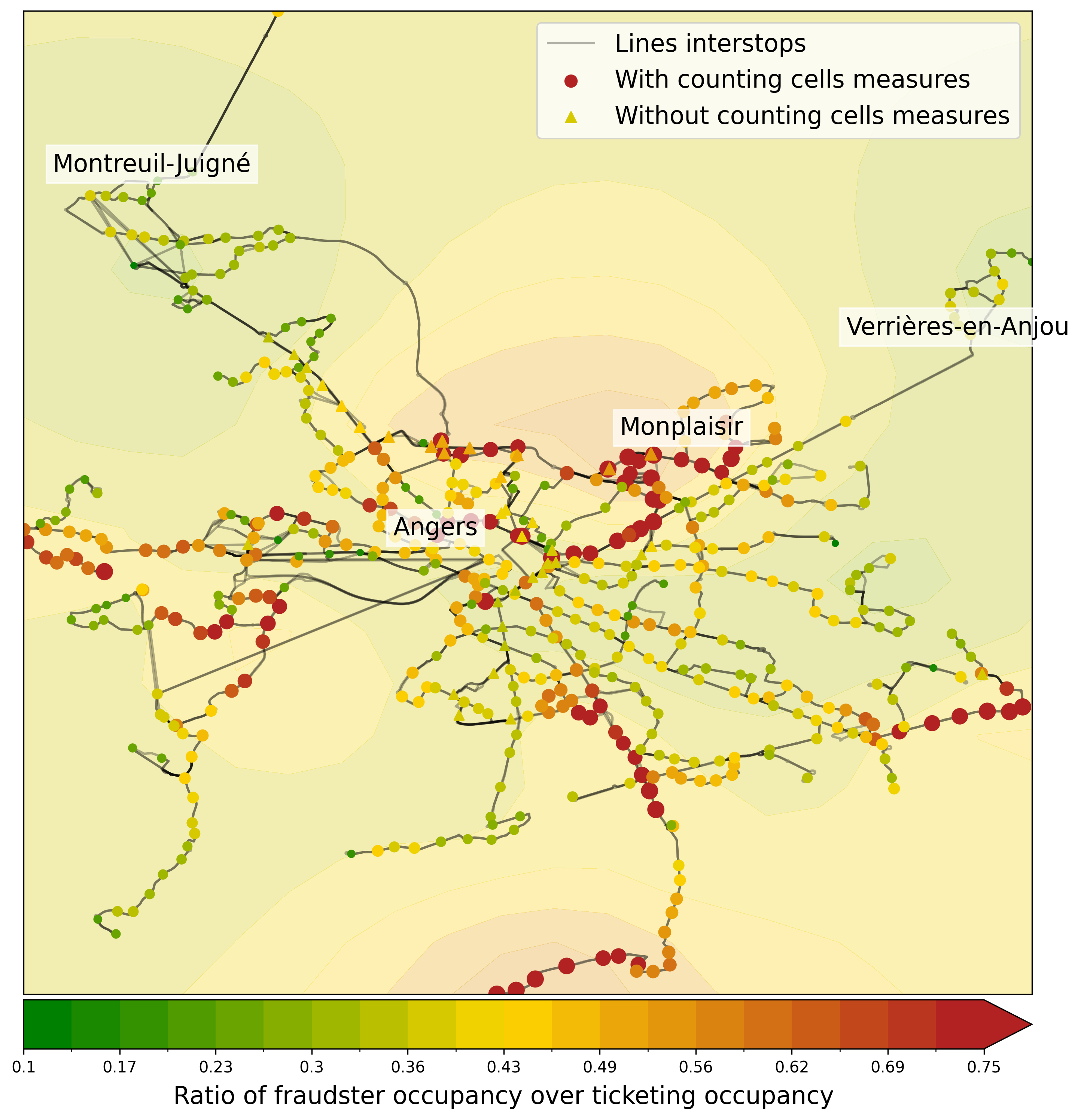}
    \caption{Fraud heatmap in the Angers area in November 2022.}
    \label{fig:fraudmap_irigo}
\end{figure}

Fig.~\ref{fig:fraudmap_irigo} illustrates the inferred fraud map on the Angers network, with stations with and without mean fraud rate highlighted. Varying fraud levels can be observed in the different regions on the map. The stations that do not have counts available (marked with triangles) are mainly located in the central district of the city.

The behaviour of fraudsters is highly cultural and therefore this analysis may not be applicable outside of France \cite{lutteFraudeMinistere,cc2016lutte}. For instance, high fraud rates are observed in the Monplaisir district, one of the Northern suburbs of Angers, which is considered the poorest district of the area \cite{inseeAngers}. In this neighborhood, fraud is likely due to economic reasons \cite{perrotta2017}, rather than other fraud justifications (incidental, contesting, etc.).
Fraud is also known to be higher at terminals as these areas tend to be less controlled. 
This is observed mainly at the southernmost section of the network, but also at the terminals of the East and West lines.

Patches of green are visible in the North-West and in the North-East: these are the suburbs of Montreuil-Juigné and of Verrières-en-Anjou, both being residential areas. Public transportation is predominantly used by schoolchildren, who are not frequent fare evaders \cite{delbosc2016cluster}.

Though fraud is common on the tramway because unlike buses the boarding of passengers can be done far from the driver \cite{BHLSreport}, this is not accurately represented here as the Angers tramways are not equipped with counting cells and are located in the city center.

\section{DISCUSSION}

While the proposed method demonstrates promising results on real-world data, it does not account for edge cases. For instance, there is no enforced minimum threshold on the number of historical courses required to compute a precise average fraud rate. Specifically, the model would generate results even with a single course that has counting cell measures over several months of data without such measures. Similarly, the geospatial regression model does not enforce a threshold on the number or topology of stations with computed fraud rates necessary to infer a reliable spatial fraud model. However, implementing such thresholds could be easily accomplished, but would require further investigations with field experts. Moreover, based on our experience, it is quite uncommon for an operator to equip only a single vehicle. Typically, counting cell equipment is part of a broader investment program where at least 10\% of vehicles are equipped on several key routes in the network. In light of the study conducted in Section~\ref{sec:cc_coverage}, we recommend operators to equip at least 10\% of their fleet with counting cells and carry out intelligent rotations to ensure measurements are available at the maximum number of stations, hours of the day and days of the week.

Regarding time range, a sufficient amount of data is necessary to compute accurate fraud rate averages. Therefore, an appropriate temporal history depth must be determined: it should be long enough to yield precise averages, but not excessively so as to prevent smoothing out seasonality changes in ridership. Given sufficient data, it may be considered to include the day of the week or the time of the day as features in calculating fraud rates.

In Section~\ref{sec:performance}, it appears that the geospatial outperforms the mean fraud rate model with the wMAPE metric. This could be attributed to a smoothing of fraud rates by the Gaussian kernels when occupancy is low but fraud is nonetheless present. However, caution is required as the metrics are computed in comparison to raw APC data and not to unobserved ground truth, notably unavailable on lines where geospatial regression applies.

\section{CONCLUSIONS}

To address the unavailability of a complete and exhaustive ridership information on public transportation networks at the course level, we have proposed a method that combines two common data sources. Our method models a fraud rate at the station level using courses with both ticketing and counting data available. This fraud rate is then applied to ticketing data (assuming its availability over the entire network) for courses without counting data. Our approach features two methods to compute the fraud rate: by averaging it at stations that have some courses stopping there with counting data or by inferring a geospatial model for the rest of the network. Additionally, this method offers a spatial visualization of fraud across the entire network coverage. The proposed algorithms have been tested and discussed on real-world data from four public transportation networks. They outperform a simpler prediction model, unable to use ticketing data nor to predict on unequipped lines. Moreover, we have analyzed the impact of counting cell coverage on the accuracy of occupancy reconstruction and interpreted an example of a fraud map that shows the consistency of the results with field knowledge and known causes of fraud.

As future work, assuming service frequency is sufficiently high, fraud rates estimation could be improved by incorporating features such as the time of the day, day of the week, and holidays. We are also working on unifying boarding and alighting counts directly instead of the occupancy. Preliminary results show that this approach reduces the error in reconstructing occupancy by mitigating imprecisions in boarding and alighting counts.

\section*{ACKNOWLEDGMENT}

We would like to thank the Irigo network (city of Angers), the Taneo network (city of Nevers) and the Bibus network (Brest) for allowing us to publish the results of our algorithms on their data. We also warmly thank the Citio team for providing us with working and computing infrastructure, clean and formatted data, and domain expert knowledge to help us analyze results.

\bibliographystyle{IEEEtran}
\bibliography{root}

\end{document}